\documentstyle[epsf,psfig]{mn}

\newif\ifAMStwofonts

\def\Romain#1{\expandafter\uppercase\expandafter{\romannumeral #1}}
\def\ion#1#2{#1$\;${\small\rm\Romain#2}\relax}

\ifoldfss
  \ifCUPmtlplainloaded \else
    \NewTextAlphabet{textbfit} {cmbxti10} {}
    \NewTextAlphabet{textbfss} {cmssbx10} {}
    \NewMathAlphabet{mathbfit} {cmbxti10} {} 
    \NewMathAlphabet{mathbfss} {cmssbx10} {} 
  \fi
  \ifAMStwofonts
    \ifCUPmtlplainloaded \else
      \NewSymbolFont{upmath} {eurm10}
      \NewSymbolFont{AMSa} {msam10}
      \NewMathSymbol{\upi}     {0}{upmath}{19}
      \NewMathSymbol{\umu}     {0}{upmath}{16}
      \NewMathSymbol{\upartial}{0}{upmath}{40}
      \NewMathSymbol{\leqslant}{3}{AMSa}{36}
      \NewMathSymbol{\geqslant}{3}{AMSa}{3E}

      \let\leq=\leqslant 
      \let\geq=\geqslant 
    \fi
  \fi
\fi 

\ifnfssone
  \newmathalphabet{\mathit}
  \addtoversion{normal}{\mathit}{cmr}{m}{it}
  \addtoversion{bold}{\mathit}{cmr}{bx}{it}
  \newmathalphabet{\mathbfit} 
  \addtoversion{normal}{\mathbfit}{cmr}{bx}{it}
  \addtoversion{bold}{\mathbfit}{cmr}{bx}{it}
  \newmathalphabet{\mathbfss} 
  \addtoversion{normal}{\mathbfss}{cmss}{bx}{n}
  \addtoversion{bold}{\mathbfss}{cmss}{bx}{n}
  \ifAMStwofonts
    \ifCUPmtlplainloaded \else
      \UseAMStwoboldmath
      \makeatletter
      \new@mathgroup\upmath@group
      \define@mathgroup\mv@normal\upmath@group{eur}{m}{n}
      \define@mathgroup\mv@bold\upmath@group{eur}{b}{n}
      \edef\UPM{\hexnumber\upmath@group}
      \new@mathgroup\amsa@group
      \define@mathgroup\mv@normal\amsa@group{msa}{m}{n}
      \define@mathgroup\mv@bold\amsa@group{msa}{m}{n}
      \edef\AMSa{\hexnumber\amsa@group}
      \makeatother
      \mathchardef\upi="0\UPM19
      \mathchardef\umu="0\UPM16
      \mathchardef\upartial="0\UPM40
      \mathchardef\leqslant="3\AMSa36
      \mathchardef\geqslant="3\AMSa3E

      \let\leq=\leqslant 
      \let\geq=\geqslant 
    \fi
  \fi
\fi 

\ifnfsstwo
  \DeclareMathAlphabet{\mathbfit}{OT1}{cmr}{bx}{it}
  \SetMathAlphabet\mathbfit{bold}{OT1}{cmr}{bx}{it}
  \DeclareMathAlphabet{\mathbfss}{OT1}{cmss}{bx}{n}
  \SetMathAlphabet\mathbfss{bold}{OT1}{cmss}{bx}{n}
  \ifAMStwofonts
    \ifCUPmtlplainloaded \else
      \DeclareSymbolFont{UPM}{U}{eur}{m}{n}
      \SetSymbolFont{UPM}{bold}{U}{eur}{b}{n}
      \DeclareSymbolFont{AMSa}{U}{msa}{m}{n}
      \DeclareMathSymbol{\upi}{0}{UPM}{"19}
      \DeclareMathSymbol{\umu}{0}{UPM}{"16}
      \DeclareMathSymbol{\upartial}{0}{UPM}{"40}
      \DeclareMathSymbol{\leqslant}{3}{AMSa}{"36}
      \DeclareMathSymbol{\geqslant}{3}{AMSa}{"3E}

      \let\leq=\leqslant 
      \let\geq=\geqslant 
    \fi
  \fi
\fi 

\ifCUPmtlplainloaded \else
  \ifAMStwofonts \else 
    \def\upi{\pi}
    \def\umu{\mu}
    \def\upartial{\partial}
  \fi
\fi

\title[ HST imaging of the CFRS and LDSS surveys IV. ]{Hubble Space 
Telescope Imaging of the CFRS and LDSS Redshift Surveys -- IV. Influence of
mergers	in the evolution of faint field galaxies from $z\sim1$}

\author[Le F\`evre et al.]{O.~Le~F\`evre,$^1$ R.~Abraham,$^2$  
S.J.~Lilly,$^3$ R.S.~Ellis,$^2$ J.~Brinchmann,$^2$ D.~Schade,$^6$ 
\newauthor
L.~Tresse,$^{1,4}$ M.~Colless,$^5$ D.~Crampton,$^6$ 
K.~Glazebrook,$^7$ F.~Hammer,$^8$  T.~Broadhurst$^9$ \\
$^1$ Laboratoire d'Astronomie Spatiale--CNRS, Traverse du Siphon, 
B.P.8, 13376 Marseille Cedex 12, France, lefevre@astrsp-mrs.fr\\
$^2$ Institute of Astronomy, University of Cambridge,  Madingley Road,
Cambridge, CB3 0HA, United Kingdom\\
$^3$ Department of Astronomy, University of Toronto, 60 St. George Street, Toronto, Ontario, M5S 3H8, Canada\\
$^4$ Istituto di Radioastronomia - CNR, Via P. Gobetti, 101, 40129 Bologna, Italy\\
$^5$ Mount Stromlo \& Siding Spring Observatories, Australian National University, Weston Creek, Canberra, ACT 2611, Australia\\
$^6$ Dominion Astrophysical Observatory, 5071 West Saanich Road,
Victoria, B.C., V8X 4M6, Canada\\
$^7$ Anglo--Australian Observatory, P.O. Box 296, Epping, NSW 1710, Australia \\
$^8$ DAEC, Observatoire de Paris--Meudon, 92195 Meudon Cedex, France\\
$^9$ Department of Astronomy, University of California, 601 Campbell Hall, University of California, Berkeley CA 94720-3411, USA }

\date{Accepted 1999 September}

\pagerange{\pageref{firstpage}--\pageref{lastpage}}
\pubyear{1999}

\begin{document}

\maketitle

\label{firstpage}

\begin{abstract}
{\it Hubble Space Telescope} images of a sample of 285 galaxies with
measured redshifts from the CFRS and Autofib--LDSS redshift surveys
are analysed to derive the evolution of the merger fraction out to
redshifts $z \sim 1$. We have performed visual and machine-based
merger identifications, as well as counts of bright pairs of
galaxies with magnitude differences $\delta m \leq 1.5$ mag. We
find that the pair fraction increases with redshift, with up to
$\sim20\%$ of the galaxies being in physical pairs at
$z\sim0.75-1$.  We derive a merger fraction varying with redshift
as $\propto (1+z)^{3.2\pm 0.6}$, after correction for
line-of-sight contamination, in excellent agreement with the
merger fraction derived from the visual classification of mergers
for which $m=3.4\pm0.6$. After correcting for seeing effects on
the ground-based selection of survey galaxies, we conclude that
the pair fraction evolves as $\propto (1+z)^{2.7\pm 0.6}$. This
implies that an average L$^{*}$ galaxy will have undergone 0.8 to 1.8
merger events from $z=1$ to $z=0$, with 0.5 to 1.2 merger events
occuring in a 2 Gyr time span at around $z\sim0.9$. This result is
consistent with predictions from semi-analytical models of galaxy
formation. From the simple co-addition of the observed
luminosities of the galaxies in pairs, physical mergers are
computed to lead to a brightening of 0.5 mag for each pair on
average, and a boost in star formation rate of a factor of 2, as
derived from the average [\ion{O}{2}] equivalent widths. Mergers of
galaxies are therefore contributing significantly to the evolution
of both the luminosity function and luminosity density of the
Universe out to $z \sim 1$.
\end{abstract}


\section{Introduction}

Mergers of galaxies have long been known to play an important role
in the evolution of galaxies. Detailed case studies in the local
Universe have shown the powerful effect mergers can have on galaxy
morphologies and on the triggering of star formation (e.g.
Kennicutt 1996; Schweizer 1996). The importance of mergers has
also been emphasized by computer simulations involving either
major mergers of galaxies with comparable mass, or the merger of
dwarf galaxies with a more massive galaxy (Mihos \& Hernquist
1994a,b; Hernquist \& Mihos 1995; Mihos 1995, hereafter M95).

If merging is at work at all during the lifetime of galaxies, then
the space density, mass, luminosity, and morphology of galaxies
must change with epoch. Several authors suggest that merging
systems can explain part (or all) of the excess number of galaxies
observed in deep photometric galaxy counts
(Rocca-Volmerange \& Guiderdoni 1990; Broadhurst, Ellis \&
Glazebrook 1992). Mergers can influence such surveys either by the
separate counting of individual systems ultimately destined to
merge, or through more subtle effects. For example, dissipative
mergers are expected to trigger some degree of star formation, so
in magnitude-limited samples one expects to observe less massive
galaxies brightened above observational limits by merger-induced
star-formation activity.

The evolution of the merger rate with redshift is a key observable
that can be used to test galaxy formation models (see Abraham 1998 for
a review). In particular, semi-analytical models for galaxy formation
in hierarchical clustering scenarios postulate that galaxies assemble
from the successive merger of smaller sub-units.  Such models make
concrete predictions for the relationships between evolution in the
morphological mix and redshift-dependent merger rates (Baugh, Cole \&
Frenk 1996; Baugh et al. 1998; Kauffmann, White \& Guiderdoni 1993;
Kauffmann 1996).

Evolution of the luminosity function of galaxies at high redshifts
(Lilly et al. 1995, Ellis et al. 1996) indicates that, on the
whole, either the typical luminosity of galaxies was brighter by
$\sim$1 magnitude at $z\sim0.7$, or that there were $\sim$3 times
more galaxies at this epoch, or that a combination of both
luminosity and number density evolution is occurring.
Disentangling the effects of these two phenomena is central to
understanding the nature of field galaxy evolution. Luminosity
evolution has been shown to play an important role in the
evolution of disk galaxies, from observations with the {\em Hubble
Space Telescope} (HST) of systems from the CFRS--LDSS sample
(Schade et al. 1995; Lilly et al. 1998). On the other hand,
number-density evolution remains poorly constrained.

There are several reasons why meaningful constraints on
number-density evolution have been difficult to obtain. One needs
large galaxy samples with redshifts measured up to large look-back
times {\em and} deep high spatial resolution images in order to
observe the close environment of galaxies at better than 1~kpc
resolution. Furthermore, the time-scales associated with merging
are not straightforward to constrain. There is considerable
uncertainty in any transformation from an ``instantaneous'' merger
rate (computed directly from observations at a given look-back
time) to a ``global'' merger rate indicating the likelihood for a
galaxy to experience merging from any epoch to the present.

Data from HST indicates that, at moderate redshifts, there is a larger
proportion of faint galaxies exhibiting peculiar morphologies
suggestive of merging (Griffiths et al. 1994; Driver et al. 1995;
Glazebrook et al. 1995, Neuschaefer et al. 1997). Following early
attempts to measure the evolution of the merger fraction (Zepf \& Koo
1989), these HST observations complement studies of close pairs of
galaxies from redshift survey samples (Carlberg et al. 1994; Yee \& 
Ellingson 1995; Woods et al. 1995; Patton et al. 1997) out to
intermediate redshifts $z\sim0.3$. In a pair study of a complete
sample of galaxies with measured redshifts, Patton et al. (1997) find
that at a mean redshift $z\sim0.33$, $4.7\pm0.9$\% of galaxies are in
close physical pairs, compared to $2.1\pm0.2$\% locally (Patton et
al. 1997), leading to a change in the merger rate with redshift
proportional to $(1+z)^{2.8\pm0.9}$.

The Canada--France Redshift Survey (CFRS, Lilly et al. 1995; 
Le~F\`evre et al. 1995), and LDSS surveys (Glazebrook et al. 1995;
Colless et al. 1993), provide  samples with several hundred
galaxies with redshifts measured up to $z\sim1.2$. Well-defined
sub-samples from these surveys have been observed with HST,
allowing us to obtain quantitative information on the
morphological type mix (Brinchmann et al. 1998), the central disk
and bulge surface brightnesses (Lilly et al. 1998; Schade et al. 
1999), and, in the present paper,  the close environment of the
survey galaxies and the merger rate as a function of redshift.

In the present paper we present the measurements of the merging
rate of galaxies out to redshifts $\sim1$, from HST cycle~4 and
cycle~5 imaging of a total sample of 285 galaxies. The paper is
organized as follows. Our sample is described in Section~2. Visual
identifications of merging galaxies, as well as pair counts based
on both visual and computer-based classifications, are presented
in Section~3. The evolution of the merger rate out to $z\sim1$ is
computed in Section~4, and the luminosity enhancement and
selection effects introduced by mergers on the magnitude limited
sample are evaluated in Section~5. A discussion of the results is
presented in Section~6. We use $H_0=50$ km~s$^{-1}$~Mpc$^{-1}$ and
$q_0=0.5$  throughout this paper.

\section{HST imaging sample}

Imaging data was primarily acquired during HST cycles 4 and 5 with
the Wide Field and Planetary Camera 2 (WF/PC2) and the F814W
filter. In addition, field CFRS~1415+52 (set on a deep VLA field;
Fomalont et al. 1991) was observed with HST as part of the ``Groth
strip'' campaign (Groth et al. 1994), and was included in the
present data set. Exposure times varied between 4400s and 7400s,
and allowed us to reach signal-to-noise ratios of S/N$=$1 per WFPC2
pixel for $\mu_{I_{AB}}=25.3$ mag~arcsec$^{-2}$. This surface
brightness limit is deep enough to trace an unevolved Milky
Way-like disk to three scale lengths at the redshift limit of our
combined survey. The full CFRS+LDSS HST imaging dataset is
described in detail in Brinchmann et al. (1998). A total of 232
CFRS galaxies and 53 Autofib+LDSS galaxies have both a securely
measured redshift in the range $0.05 \leq z \leq 1.2$ (with a
redshift identification secure at the 85\% level: Le~F\`evre et
al. 1995; Brinchmann et al. 1998, Table~II), and a deep HST/WFPC2
image. Galaxies in the CFRS  are drawn from a purely $I$-band
selected sample, as defined in Lilly et al. (1995) and Le F\`evre
et al. (1995). Galaxies in the LDSS sample are a purely
$B$-selected sample, as defined in Colless et al. (1993) and
Glazebrook et al. (1995). Galaxies near the edge of the WFPC2
field (those systems for which a full $20 h^{-1}$ kpc area around
them was not visible) were removed from the sample. The redshift
and magnitude distributions of the HST samples are similar to the
distributions of their parent CFRS and LDSS samples (as described
by Brinchmann et al. 1998).

\section{Image Analysis}

\subsection{Visual identification of mergers}\label{vim}

In this section we seek to identify merger events through visual
classification. In this procedure we attempt to distinguish
broadly between probable ``major'' and ``minor'' merging events. A
major merger, involving galaxies of comparable masses, should
produce strong morphological signatures, such as double nuclei,
wisps, tails, shells, asymmetric morphology, and possibly be
associated with the triggering of a recent starburst (Mihos \&
Hernquist 1994a; Liu \& Kennicutt 1995).  Minor mergers, involving
a massive galaxy and a small dwarf, are likely to be more frequent
(M95; Baugh, Cole \& Frenk 1996), and produce more benign morphological
signatures which might be difficult to identify during or shortly
after the merging event, although significant changes of the star
formation rate can result from this type of event. It is
emphasized that the visibility of merger-induced features may be
relatively short lived, depending on the respective masses of the
galaxies involved (M95), making the identification of even
relatively recently completed mergers a challenging task.


We have defined the following visual classification scheme to
identify merger events:\\ (0) not a merger: no trace of merger,
although some minor asymmetry may be present\\ (1) suspicious:
there is a suspicion that some merging event might be occurring,
with some asymmetry and traces of wisps, tails, etc.\\ (2)
probable merger / upcoming merger: there is a nearby galaxy but
the isophotes are not overlapping at the $\mu_{I_{AB}}=25.3$
mag~arcsec$^{-2}$  isophote level.\\ (3) on-going merger: a merging
event is most probably or certainly occurring, as evidenced by
strong asymmetry, double nuclei, prominent wisps, tails, etc.

In the following analysis, classes 2 and 3 are summed to produce the
list of visually identified mergers. The separation in classes 2 and 3
is used in Section~\ref{OII}, but we emphasize that the separation
between classes 2 and 3 is sensitive to the depth of the limiting
isophote chosen, as the primary and companion galaxy may well be
``bridged'' by luminous material at levels fainter than the
observations are able to reveal.

\begin{figure}
\begin{center}
\leavevmode
\psfig{figure=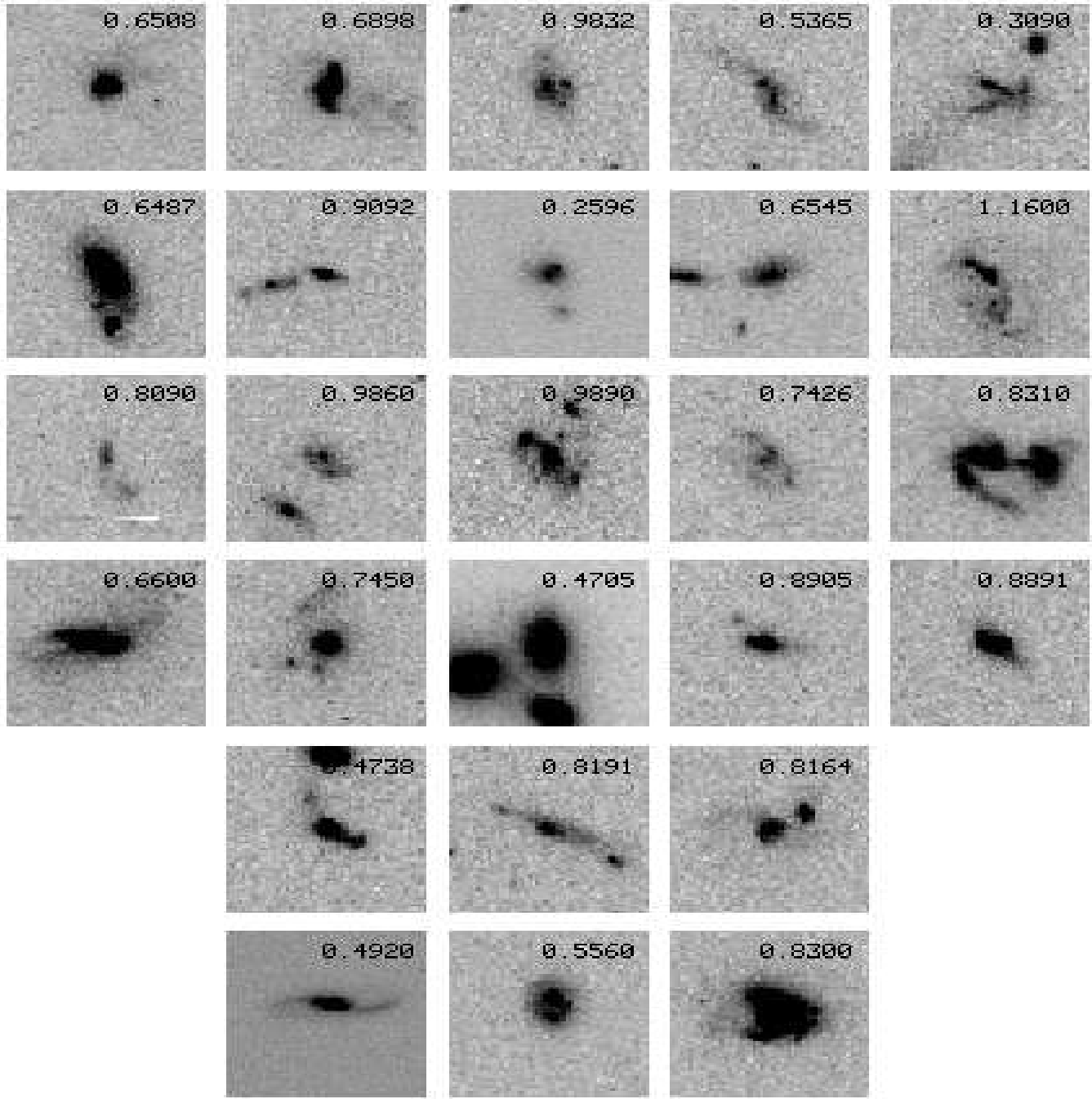,height=18cm,width=18cm}
\caption{ Galaxies visually classified as mergers. Each individual image is
$5\times5$ arcsec$^2$, the measured redshift is indicated on the upper
right corner of each image.} \label{fig:imerg}
\end{center}
\end{figure}

\begin{figure}
\begin{center}
\leavevmode
\psfig{figure=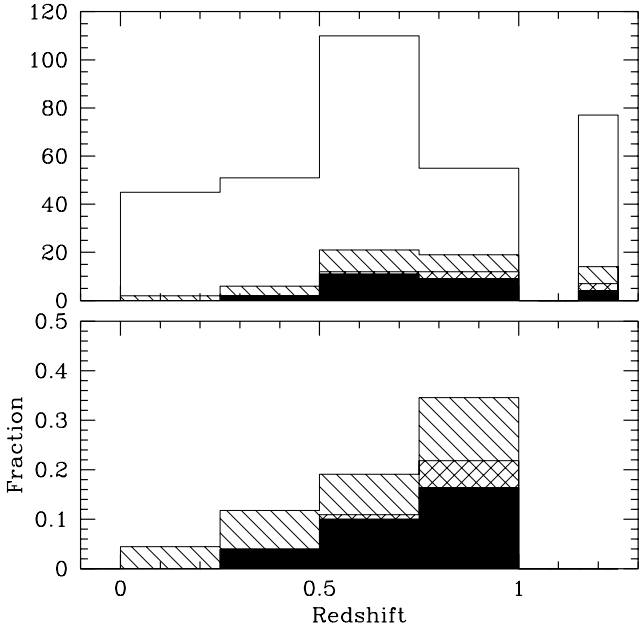,height=9cm,width=9cm}
\caption{ Numbers ({\it top}) and fraction of galaxies ({\it bottom})
visually classified as mergers. The on-going mergers (class 3, see
text) are identified as the filled histogram, probable mergers are
identified as the double crossed histogram (class 2), and suspicious
galaxies are identified by the single crossed histogram (class 1). The
top panel shows as a separate histogram (right) the distribution in
classes for the galaxies without redshift.}  \label{fig:pmerg}
\end{center} 
\end{figure}

All galaxies in the sample have been visually examined by 3 of us (JB,
OLF, SJL), and classified in the above scheme. When a classification
was discrepant among the 3 observers, the median class has been taken
as the final classification. A total of 261 galaxies have a secure
redshift measurement; among these, 22 are classified in class 3, and 4
in class 2. This leads to an overall merger fraction for the whole
sample of $10\pm2$\%. A total of 63 galaxies do not have a redshift,
of which 3 are in class 2 and 4 are in class 3, giving a merger
fraction for this sample of $11\pm4$\%. Galaxies without redshifts
therefore have a merger fraction comparable to that for the whole
sample, and no bias in the merger rate estimate is expected from
restricting the analysis to galaxies with secure redshifts. The list
of visually identified mergers is given in Table~1. Images of the
mergers are presented in Fig.~\ref{fig:imerg} and the histogram of the
redshift distribution is shown in Fig.~\ref{fig:pmerg}. The fraction
of visually identified mergers vs. redshift is shown in
Fig.~\ref{fig:pmerg}: it is evident that most of the identified
mergers are at the high redshift end of the sample.

\subsection{Pair fraction}

Another approach to identifying mergers is to count pairs of
galaxies as observed in projection, and to use these pair counts
to  estimate the physical pair fraction in the sample (see e.g.
Patton et al. 1997). Physical pairs of galaxies separated by
small distances and with low relative speeds are bound and are
destined to merge on relatively short timescales. We have
performed a pair count analysis using both a visual approach
(based on counts of galaxies with measured redshifts for which
nearby galaxies form a pair as observed in projection), and using
the machine-based automated approach described in Section~\ref{slee}.
Because we usually have knowledge of the redshift for only one of
the pair components, some of the pairs can be produced by
foreground-background projection contamination. This effect needs
to be corrected for in order to estimate physical pair counts,
which can then be related to the merger rate. This section
describes the manual and automated classification methods used to
estimate the apparent pair fraction, leading to the computation of
the physical pair fraction in the next Section.

\begin{table}
\begin{center}
\caption{ List of visually identified mergers. 
$I_{AB}$ and $M_B(AB)$ are the I-band and
absolute blue magnitudes in the AB reference system;
EW([\ion{O}{2}]) is the rest-frame [\ion{O}{2}]3727 equivalent width.
See Section~\ref{vim} for the definition of the merger class.}
\begin{tabular}{|l|c|c|c|c|c|}
\hline
        Num & Redshift  & $I_{AB}$ &  $M_B(AB) $ &  Rest  & Merger \\
            &   z  & Ground   &        &  EW([\ion{O}{2}]) & Class \\ \hline
03.0523  & 0.6508 & 21.31 & $-$21.19 & 30.3 & 3 \\
03.1540  & 0.6898 & 21.04 & $-$21.51 & 11.8 & 3 \\
10.0761  & 0.9832 & 22.07 & $-$21.55 & 8.1  & 3 \\
10.0765  & 0.5365 & 22.18 & $-$19.95 & 52.7 & 3 \\
10.0802  & 0.3090 & 21.70 & $-$19.16 & 32.9 & 3 \\
10.1183  & 0.6487 & 20.60 & $-$21.92 & 34.6 & 3 \\
10.1220  & 0.9092 & 22.36 & $-$20.95 & 20.4 & 2 \\
14.0377  & 0.2596 & 20.81 & $-$19.61 & 62.7 & 3 \\
14.0485  & 0.6545 & 22.16 & $-$22.44 & 35.7 & 2 \\
14.0547  & 1.1600 & 21.40 & $-$23.09 & 6.0  & 3 \\
14.0665  & 0.8090 & 22.97 & $-$20.78 & 12.2 & 2 \\
14.0743  & 0.9860 & 22.19 & $-$22.05 & 14.1 & 2 \\
14.0846  & 0.9890 & 21.81 & $-$21.91 & 40.9 & 3 \\
14.1126  & 0.7426 & 22.26 & $-$20.64 & 41.9 & 3 \\
14.1129  & 0.8310 & 21.12 & $-$22.10 & 15.3 & 3 \\
14.1139  & 0.6600 & 20.20 & $-$22.27 & 12.0 & 3 \\
14.1415  & 0.7450 & 21.06 & $-$21.62 & 0.  & 3 \\
22.0497  & 0.4705 & 18.42 & $-$22.93 & 0.   & 3 \\
22.0576  & 0.8905 & 22.29 & $-$20.97 & 73.5 & 3 \\
22.0599  & 0.8891 & 21.74 & $-$21.51 & 68.8 & 3 \\
22.0919  & 0.4738 & 21.77 & $-$20.23 & 10.2 & 3 \\
22.1313  & 0.8191 & 21.74 & $-$21.33 & 57.2 & 3 \\
22.1453  & 0.8164 & 21.44 & $-$21.53 & 0.   & 3 \\
10.12073  & 0.4920 & 20.35 & $-$20.61 & 0.  & 3 \\
13.12106  & 0.5560 & 21.70 & $-$18.68 & 9.0 & 3 \\
13.12545  & 0.8300 & 20.30 & $-$21.26 & 24.0 & 3 \\
 \\ \hline
\end{tabular}
\end{center}
\end{table}

\subsubsection{Visual identification of galaxy pairs}\label{s2}

We have determined pair counts based on galaxies with at least one
nearby companion galaxy within a $20 h^{-1}$ kpc radius, and a
magnitude difference $\delta m \leq 1.5$ mag between the main
galaxy and the companion. Such systems are expected to merge
within less than $\sim10^9$ years (M95; Patton et al. 1997). The
difference in magnitude is intended to identify unambigous pairs,
and to produce pair counts less affected by the difficulties
inherent in separating minor mergers of dwarf galaxies with bright
galaxies from galaxies with large \ion{H}{2} regions or asymmetric
features.  
\onecolumn
\twocolumn
\noindent
These minor mergers can certainly contribute to the
general evolution of the morphology and luminosity of galaxies,
but the corresponding pair counts are subject to larger
uncertainties than the bright pair counts, because of possible
significant background contamination. The $\Delta m \leq 1.5$ mag
cut-off therefore leads to an underestimate of the true pair
fraction.

The visual measurements were carried out as follows. For each
galaxy in the CFRS-LDSS catalog with an HST/WFPC2 image, we
measured the projected distances
$d_1,...,d_i$ to putative companion galaxies, and their
corresponding magnitude differences $\delta m_1,...,\delta m_i$
with respect to the primary galaxy. Objects sharing a common
isophote with the primary galaxy were noted in an attempt to
broadly distinguish between ``upcoming'' and ``on-going'' mergers
(see Section~\ref{OII}). Galaxy pairs were flagged as ``1'' when
the separation
$$d_{\theta}= \frac{c}{H_0} \frac{q_0z + (q_0-1)\times (-1+(1+2q_0z)^{0.5}}
{q_0^2 \times (1+z)^2} $$
between the primary galaxy and the companion was less
than $20 h^{-1}$ kpc, and the magnitude difference $\delta m_i$
was less than 1.5 magnitude.

\subsubsection{Automated identification of pairs: Lee classifier}\label{slee}

In addition to the visual approach defined above, we have
undertaken the measurement of a quantitative parameter indicative
of merging. The degree of bimodality in the galaxy images is
determined objectively from measurements of the Lee Ratio (Abraham
et al. 1998), $L_R$, for each of the galaxy images. Our definition
of $L_R$ is a straightforward generalization of the
maximum-likelihood statistic developed by Fitchett \& Webster(1987) for use
with sparsely sampled galaxy cluster data (see also
Lee, 1979). Sky-subtracted 100x100
pixel ``postage stamp'' images were constructed for each galaxy in
the sample, and the galaxian light isolated from the sky
background at the 2$\sigma$ level. The galaxian light distribution
$I_{i,j}$ at pixel $i,j$ in each of these images was then
projected onto a line at angle $\theta$, defining a profile
$P(\,r\mid\theta\,)$, where $r$ is the distance along the axis of
projection:
$$P(\,r\mid\theta\,) = \sum_{i=-50}^{50} \sum_{j=-50}^{50}
             I_{i,j}\,\delta(r - i \cos\theta - j \sin\theta). $$
For each projection, a function $L$ was determined by
subdividing the projection at a number of cuts:
$$L(\,d\mid\theta\,) = {(\mu_l - \mu)^2 + (\mu_r - \mu)^2 \over
                  \sigma_l + \sigma_r},$$
where $d$ is the position of the cut, $\mu_l$ and $\sigma_l$
[$\mu_r$ and $\sigma_r$] are the mean and standard deviations of the
profile to the left [right] of the cut, and $\mu$ is the mean of the
original profile, i.e.
$$\mu_l(\,d\mid\theta\,) = \sum_{w=-50}^{d} w P(w | \theta)$$
$$\mu_r(\,d\mid\theta\,) = \sum_{w=d}^{50} w P(w | \theta)$$
$$\mu(\theta) = \sum_{w=-50}^{50} w P(w | \theta)$$
$$\sigma_l(d\mid\theta) = \sum_{w=-50}^{d} (w - \mu_l)^2 P(w |
\theta)$$ $$\sigma_r(d\mid\theta) = \sum_{w=d}^{50} (w - \mu_r)^2
P(w | \theta).$$
The Lee Ratio $L_R$ is then defined as the ratio of the
maximum to the minimum of the $L$ for all angles and all cuts:
$$L_R = {\max_{0< \theta <\pi\atop-50< d <50 } L(\,d\mid\theta\,)
         \over
         \min_{0< \theta <\pi\atop-50< d <50 } L(\,d\mid\theta\,)}.$$
$L_R$ was calibrated using images of synthetic
``mergers'' constructed by adding together distinct galaxy images
with a range of relative brightnesses and displaced across a range
of separations spanning the range observed in the HST image sample
(Abraham et al. 1998). $L_R$ was computed in steps of 5 degrees.
$L_R$ is most sensitive to cases where the magnitude difference
between two galaxies is small, and where the isophotes between
primary and companion galaxies are overlapping. The $L_R$
parameter is therefore sensitive only to {\it major} mergers.
Values of $L_R > 1.5$ are indicative of significant bimodality.

\subsubsection{Comparison of manual and automated classifications}\label{s1}


A comparison between manually and automatically classified ($L_R
\geq 1.5$) pairs of galaxy images was made. Of the 49 pairs
identified by the manual approach,  37(80\%) have a Lee ratio $L_R
\geq 1.5$. The remaining 12 galaxies were examined in detail, and
it was found that of these the manual classification had picked
out 10 systems with well defined bimodal components, while  2
galaxies with low $L_R$ were wrongly classified as mergers during
the manual classification. The majority of bimodal systems not
flagged as such by $L_R$ were found to have been rejected because
of the 2$\sigma$ isophote cut imposed on the data fed into the Lee
classification program -- at this level a few weakly bridged
systems were split into individual objects which separately have
low Lee ratios. Lowering the isophote limit to the level used for
the manual classification in Section~\ref{s2} recovered most of
the same pairs identified in the manual classification. In
addition, a total of 37 galaxy images (of 285) have $L_R \geq 1.5$
but were not classified as major mergers through the visual
classification scheme. Most of these were not classified as pairs
either because the magnitude difference between the components of
the pairs was $\delta m \geq 1.5$, or because there was a
companion galaxy, but outside of the $r=20 h^{-1}$ kpc area, or
because a bright unresolved object (presumably a star) was nearby.
The manual and the $L_R$ classifications are therefore broadly
consistent.

\begin{figure}
\begin{center}
\leavevmode
\psfig{figure=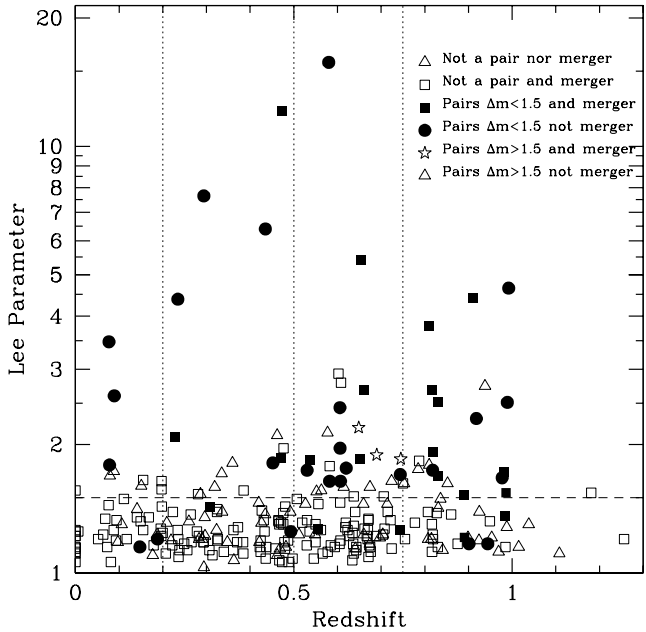,height=9cm,width=9cm} 
\caption{ Lee paremeter vs. redshift. Symbols used for
the different categories of galaxies are identified in the upper
right corner.} \label{fig:plee}
\end{center}
\end{figure}

In Fig.~\ref{fig:plee}, the $L_R$ parameter for all galaxies in
the sample is presented, and the 49 galaxies manually classified
as pairs are identified. This comparison leads to the final list
of pairs reported in Table~2, and to the images shown in
Fig.~\ref{fig:pm11}.

\subsubsection{Comparison between visually identified
mergers and galaxy pairs}

Among galaxies visually classified as mergers, 21 galaxies are
also classified as pairs, and 5 galaxies are not (galaxy numbers
10.1183, 14.377, 14.0547, 14.1415, 10.12073). These 5 galaxies
exhibit signs of asymmetry, faint wisps, or tails -- leading to
the merger classification -- but without double nuclei or close
companions. The remaining systems all exhibit double nuclei or
close companions, and warrant classification as pairs.

\subsubsection{Physical pair fraction}

The identification of pairs in our images is based on projected
appearance. Since in essentially all cases only the redshift of
the brightest galaxy is available from our ground-based surveys,
we need to correct the number of observed pairs for the appearance
of non-physical pairs originating in background/foreground
contamination along the line of sight, before we can obtain an
estimate of the true number of physical pairs. The
background/foreground contamination correction was calculated
using the counts published by Driver et al. (1995) and Abraham et
al. (1996), integrated over the magnitude range
$m_{main}<m_{gal}<m_{main}+1.5$ and within a projected $20 h^{-1}$
kpc radius area. 
At $z\sim1$, the limiting magnitude $I_{AB}=24.5$ of the HST
images allows us to securely identify all galaxies brighter than
$M_B=-17.5$. For each galaxy, this $M_B$ was then used with the
K-correction, $k(z)$, of the primary galaxy to compute the apparent
magnitude down to which the counts were to be performed, and the
background/foreground contamination applied. It is noteworthy that
the detailed form of the K-correction has only a minor effect on
this calculation, since observed $I$-band corresponds to rest
$B$-band at $z \sim 0.9$.

Our results are summarized in Table~3, which lists the numbers of
galaxies, observed pairs, and expected line of sight
contamination. The pair fraction reported in Table~3 has been
computed as the fraction of pairs in the sample, corrected for the
expected number of non-physical projected pairs. The statistics of
our survey are such that the pair fraction can be measured with
good precision for the highest two redshift bins, while at $z<0.5$
the numbers of pairs are smaller and the background/foreground
contamination is higher because $20 h^{-1}$ kpc projects to a
larger projected angular area.

\section{Evolution of the merger rate}\label{s4}

Our data allow us to determine the first direct observational
measurement of the merger fraction at redshifts $z>0.5$. We can
derive the merger fraction from both (complementary) approaches
that we have followed. Within the current sample, both the
fraction of mergers and the pair fraction rise by a factor of more
than 2 between redshifts 0.5 and 0.9. Compared to local
measurements of the merger fraction (Patton et al. 1997), the
merger fraction at z$\sim$0.9 is more than 9 times higher than in
the local universe.

From the visual classification of mergers, we find that the merger
fraction is $5.8\pm2.6$\%, $10.5\pm3.5$\%, $21.8\pm6.3$\%, at mean
redshifts 0.37, 0.63 and 0.9 respectively. If we take the local
merger value to be $2.1\pm0.2$\% (Patton et al. 1997), a
least-squares fit of this data gives, for the evolution of the
merger fraction parameterized as $R_{merg}(z)= R_{merg}(0) \times
(1+z)^{m}$, $R_{merg}(0)=0.021 \pm 0.004$ and $m=3.4 \pm 0.6$. Our
redshift $z=0.37$ result is compatible with the merging fraction of
$4.7\pm0.9\%$ at $z=0.33$ from Patton et al. (1997).

\begin{figure}
\begin{center}
\leavevmode
\psfig{figure=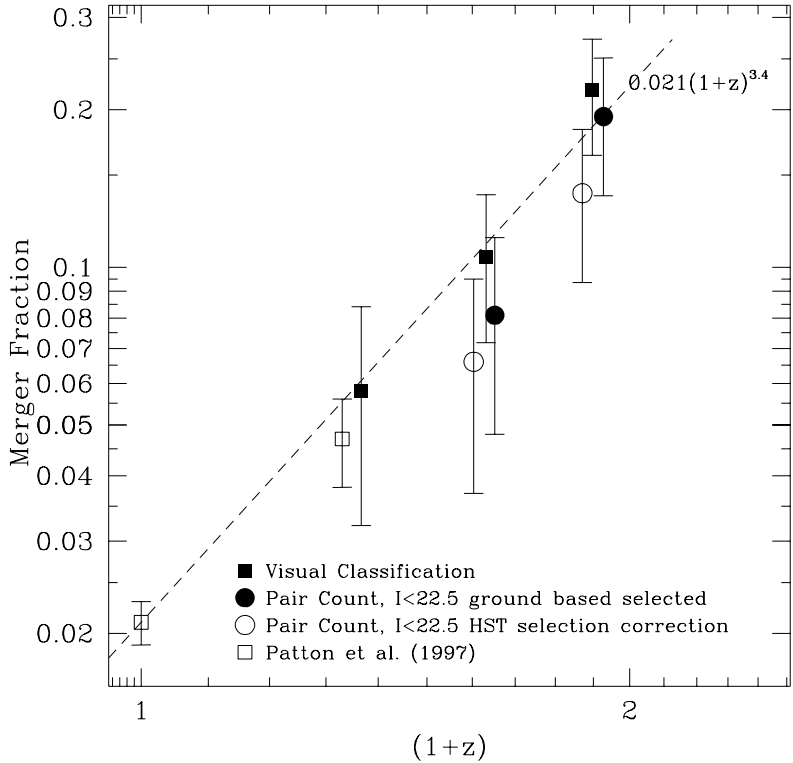,height=11cm,width=11cm} 
\caption{ Evolution of the fraction of visually confirmed mergers
(filled squares) and of the physical pair fraction (filled circles).
The physical pair fraction corrected from the smearing effect present
in a ground-based magnitude selected sample is presented as open
circles (see Section~\ref{s3}). The empty squares are from Patton et
al. (1997).  The different points at $z=0.63$ and $z=0.91$ have been
slightly shifted in the plot for clarity.}  \label{fig:pra}
\end{center} 
\end{figure}

Our pair counts (corrected for background/foreground
contamination) indicate that at redshifts $z=0.63$ and $z=0.91$,
$9.9 \pm 3.5$\% and $20.3 \pm 5.7$\% of the galaxy pairs are
likely to be physical, respectively. However, as pointed out by
Carlberg et al. (1994) and Patton et al. (1997), one needs to
determine the fraction of these physical pairs which are likely to
merge. Patton et al. (1997) suggest that, at the present epoch,
50\% of the galaxy pairs with relative velocities less than 350
km~s$^{-1}$ are likely to merge. This is expected to change with redshift
as $(1+z)^{-1}$. Applying this correction, we find that the merger
fractions at $z=0.63$ and $z=0.91$ are $8.1 \pm 3.3$\% and $19.4
\pm 5.7$\%, respectively. Using  all of the 4 CFRS--LDSS HST data
points in Table~3 in combination with the lower redshift
measurement of  $4.7 \pm 0.9$\% at $z=0.33$, and a local merging
fraction of $2.1 \pm 0.2$\% (Patton et al. 1997), we find that the
pair-count-estimated merger fraction evolves as
$(1+z)^{3.2\pm0.6}$, in excellent agreement with the merger
fraction from the morphological criteria. The data points are
represented in Fig.~\ref{fig:pra}. To minimize the impact of the
apparent magnitude selection on our sample, which selects
predominantly low luminosity galaxies at low redshifts and high
luminosity galaxies at high redshifts, we have repeated the above
analysis, limiting our sample to galaxies with $M_{B_{AB}} \geq
20.5$. We find that at redshifts $z=0.65$ and $z=0.906$, $7.8\pm
4.1$\% and $21.0\pm 6.1$\% of the galaxy pairs are physical pairs
likely to merge. Using our 4 data points and the data from Patton
et al. (1997), we find that the best fit is $R_{merg}(0)= 0.019 \pm
0.004$ and $m=3.25 \pm 0.63$.

Therefore, using both visually identified mergers and pair counts,
we conclude that the observed merger rate evolves with redshifts as
$(1+z)^m$ with {\it m} ranging from 3.2 to 3.4, over the redshift
range 0--1.

\section{Luminosity enhancement from mergers}

Mergers can change the luminosity function of galaxies through a
combination of both number density and luminosity evolution. In
addition to the obvious effects of density evolution, mergers may
also brighten galaxies to the point where systems that would
otherwise drop below the magnitude limits of a survey become
included in the sample.

Two effects can contribute to the luminosity enhancement of a
galaxy in a merging pair. In a dissipationless merger,the
luminosities of the two galaxies add with a final luminosity being
(at most) twice that of the most luminous pre-merging galaxy.
However, merging may also trigger additional stellar formation,
particularly during later stages of the merging process (Hernquist
\& Mihos 1995), increasing the total luminosity by adding to the
overall stellar content of the combined system. As star-formation
activity should have a longer timescale than the timescale during
which the morphological evidence for a merger remains (M95), a
global star formation increase might be expected in a galaxy
population subjected to a higher rate of merging. These two
effects are quantified in the following sections.

\subsection{Luminosity brightening}\label{le}

The brightening of the galaxy population induced by merging can be
one of the contributors to the observed evolution of the
luminosity function (Lilly et al. 1995; Ellis et al. 1996). To
estimate the brightening of the population from merging, we looked
at the brightest pairs selected from our $\delta m \leq 1.5$ mag
criterion. The minimum luminosity enhancement induced by a close
pair, if destined to merge, has simply been taken as the
co-addition of the luminosities of the galaxies in a pair, assuming
no extra star formation triggered by the event. We have computed
the magnitudes of each galaxy in a pair from the HST images
(excluding close companions) and a conservative estimate of the
final magnitude of the merged pair was obtained by simply adding
the luminosities from the two galaxies, the resulting magnitude
being equal to the ground based magnitude (Brinchmann et al. 
1998). The luminosity enhancement estimated by this simple
procedure is shown in column 8 of Table~2 for each pair, and is on
average 0.5 mag for the sample of galaxies classified as pairs.
There are 37 pairs above $z=0.5$, of which 19 are expected to be
the result of projection effects. This indicates that 18/151
galaxies, or $\sim 11.9 \pm 2.8$\% of the galaxies at $z \geq 0.5$
will experience a brightening of at least 0.5 magnitude within the
next billion years or so due to merging. As the merger timescales
are subject to considerable uncertainties, we can only speculate
that this fraction of galaxies is the minimum  fraction of
galaxies in the sample which would experience a brightening over a
time span of $\sim2$Gyr. We note that some of these galaxies will
be brought artificially into a magnitude limited sample produced
from ground based observations, because the total magnitude of the
galaxy is coming from closely paired galaxies as the result of
projection effects (Table~3), with each of the individual galaxies
in the pair being fainter than the magnitude limit of the sample,
as described in Section~\ref{s3}.

\subsection{Star formation rate increase}\label{OII}

To estimate the star formation enhancement produced during the
merger process, we have looked at the [\ion{O}{2}]$\lambda$3727 equivalent
width for three sub-samples: on-going mergers, upcoming mergers,
and galaxies with no companions, as classified during the analysis
detailed in Section~\ref{s2}. {\it On-going major mergers}, have
been defined as systems with evidence for double nuclei within $20
h^{-1}$ kpc, each galaxy having comparable luminosity ($\leq$1.5
mag), sometimes associated with disrupted morphologies. {\it
Up-coming} major mergers are defined as systems with a galaxy with
$\delta m \leq 1.5$ mag at a distance $d \leq 20h^{-1}$ kpc from a
galaxy in the redshift sample, but where the isophotes of the two
galaxies are not overlapping at the depth of our images. Of course
a single galaxy can have a total number of upcoming or ongoing
mergers larger than one. Differences in the [\ion{O}{2}] equivalent width
between the on-going and upcoming merger classes could then
indicate the effect of merging on the star formation rate.

\onecolumn
\begin{table}
\begin{center}
\caption{ List of identified Pairs. $L_R$ is the Lee ratio, A the
asymmetry of the light distribution as defined in Abraham et al. 
(1998), $\delta m$ and $d$ are the magnitude difference and the distance
in arcseconds and kpc between the two galaxies in the pair
respectively. A value ``9999'' indicates that $L_R$ or A could not be
computed. }
\begin{tabular}{|l|c|c|c|c|c|c|c|c|}
\hline
        Num &       z  & $I_{AB}$ &  $M_{AB}B$ &  $L_R$ &  A & Rest  &  $\delta m$ & d    \\
            &          & Ground &        &         &  & EW([\ion{O}{2}])  &
& "/kpc  \\
 (1) & (2) & (3) & (4) & (5) & (6) & (7) & (8) & (9) \\ \hline
     3.0332 &  0.1880  &  21.88 &  $-$17.63  &  1.2  &  0.323 & 6.7   &  1.39 & 0.55/2.2 \\
     3.0466 &  0.5304  &  22.47 &  $-$19.60  &  1.74 &  0.208 & 20.3  &  0.65 & 1.95/14.2  \\
     3.0485 &  0.6056  &  22.22 &  $-$20.24  &  2.44 &  9999  & 66.6  &  0.80 & 0.40/3.1  \\
     3.0488 &  0.6069  &  21.58 &  $-$20.85  &  1.64 &  9999  & 67.8  &  0.50 & 0.56/4.3  \\
     3.0523 &  0.6508  &  21.31 &  $-$21.19  &  1.85 &  0.362 & 30.3  &  1.38 & 0.35/2.7  \\
     3.0595 &  0.6061  &  21.46 &  $-$20.78  &  1.96 &  9999  & 17.4  &  1.40 & 0.15/1.1  \\
     3.1056 &  0.9440  &  22.33 &  $-$21.10  &  1.17 &  0.23  & 46.3  &  1.20 & 2.3/19.5  \\
     3.1319 &  0.6200  &  21.51 &  $-$20.67  &  1.76 &  0.071 & 25,9  &  1.50 & 0.50/3.9   \\
     3.1540 &  0.6898  &  21.04 &  $-$21.51  &  1.89 &  0.133 & 11.8  &  1.45 & 1.81/14.4  \\
    10.0761 &  0.9832  &  22.07 &  $-$21.55  &  1.36 &  0.005 & 8.07  &  1.40 & 0.44/3.7  \\
    10.0765 &  0.5365  &  22.18 &  $-$19.95  &  1.84 &  0.128 & 52.7  &  0.23 & 0.61/4.5  \\
    10.0794 &  0.5800  &  21.55 &  $-$20.53  &  15.73 &  0.212 & 0.   &  0.24 & 1.50/11.3  \\
    10.0802 &  0.3090  &  21.70 &  $-$19.16  &  1.43 &  0.172  & 32.9  &  1.00 & 1.8/10.1  \\
    10.0805 &  0.1475  &  21.45 &  $-$17.51  &  1.15 &  9999  & 0.    & -0.08 & 4.4/14.8  \\
    10.1168 &  1.1592  &  22.22 &  $-$21.99  &  9999 &  9999  & 103.3 &  0.10 & 0.77/6.6  \\
    10.1220 &  0.9092  &  22.36 &  $-$20.95  &  4.40 &  0.09  & 20.43 &  0.46 &  1.43/12.0  \\
    10.1643 &  0.2345  &  20.77 &  $-$18.93  &  4.38 &  0.177 & 0.   & -0.81 & 3.84/18.1  \\
    10.1644 &  0.0767  &  19.56 &  $-$17.24  &  3.48 &  0.220 & 0.05  & 0.81 & 3.84/7.6  \\
    14.0485 &  0.6540  &  22.16 &  $-$20.44  &  5.42 &  0.089 & 35.7  & 0.35 & 2.3/18.0  \\
    14.0600 &  1.0385  &  21.53 &  $-$22.16  &  9999 &  9999  & 26.5  & 1.26 & 0.21/1.8   \\
    14.0665 &  0.8090  &  22.41 &  $-$20.65  &  3.78 &  0.233 & -9   & 0.36 & 1.12/9.2  \\
    14.0725 &  0.5820  &  22.32 &  $-$19.76  &  1.64 &  0.213  & 31.0 &  1.50 & 0.55/4.1   \\
    14.0743 &  0.9860  &  21.65 &  $-$21.93  &  1.54 &  0.131  & -9   & -0.14 & 1.83/15.6  \\
    14.0749 &  0.8180  &  22.45 &  $-$20.53  &  1.74 &  0.207  & -9   & 0.33 & 0.39/3.2  \\
    14.0846 &  0.9820  &  21.81 &  $-$21.79  &  1.72 &  0.276  & 40.9  & 1.49  & 0.85/7.2  \\
    14.0854 &  0.9920  &  21.50 &  $-$22.21  &  4.65 &  0.06  & 0.   & 1.31 & 2.18/18.1  \\
    14.0939 &  0.9180  &  21.20 &  $-$22.15  &  2.30 &  0.219  & -9   & 0.49 & 1.35/11.4  \\
    14.1079 &  0.9011  &  21.95 &  $-$21.35  &  1.17 &  0.106 & 25.2  & 0.68  & 0.40/3.4  \\
    14.1126 &  0.7426  &  22.26 &  $-$20.64  &  1.26 &  0.109 & 41.9  & 0.95  & 0.20/1.6  \\
    14.1129 &  0.8310  &  21.05 &  $-$22.01  &  2.52 &  0.310 & -9   & 0.26 & 0.81/6.7  \\
    14.1139 &  0.6600  &  20.20 &  $-$22.27  &  2.68 &  0.279 & 12.0  & 0.62 &  0.55/4.3  \\
    14.1146 &  0.7437  &  21.72 &  $-$21.06  &  1.70 &  0.210 & 39.0  & 1.32  & 0.51/4.1  \\
    14.1193 &  0.0780  &  21.48 &  $-$16.66  &  1.79 &  0.279 & 0.   & 0.50 & 1.20/2.4  \\
    14.1264 &  0.7030  &  22.90 &  $-$19.96  &  9999 &  9999 & -9   & 0.02 & 2.02/16.1  \\
    14.1501 &  0.9890  &  21.74 &  $-$21.85  &  2.51 &  0.324 & -9   & 0.03 & 0.65/5.5  \\
    22.0497 &  0.4700  &  18.42 &  $-$22.93  &  1.86 &  9999 & 0.   & -0.25 & 2.16/14.9  \\
    22.0576 &  0.8910  &  22.29 &  $-$20.97  &  1.21 &  0.302 & 73.5  & 0.50 & 0.50/4.2  \\
    22.0585 &  0.2940  &  20.59 &  $-$19.46  &  7.65 &  0.053 & 0.   & 0.45 & 2.41/13.1  \\
    22.0599 &  0.8890  &  21.74 &  $-$21.51  &  1.52 &  0.439 & 68.8  & 0.10 & 0.10/0.8  \\
    22.0919 &  0.4740  &  21.77 &  $-$20.23  &  12.08 &  0.268 & 10.2  & 0.07 & 2.25/15.7  \\
    22.0953 &  0.9770  &  22.27 &  $-$21.22  &  1.67 &  0.063 & 47.5  & 1.30 & 0.70/6.0  \\
    22.1313 &  0.8190  &  21.74 &  $-$21.33  &  1.92 &  0.202 & 57.2  & 1.47 & 1.97/16.4   \\
    22.1453 &  0.8160  &  21.44 &  $-$21.53  &  2.68 &  0.076 & 0.   &  0.34 & 1.08/9.0  \\
   10.12525 &  0.4350  &  20.00 &  $-$21.35  &  6.40 &  9999 & 2.1  & 0.67 & 2.55/17.1  \\
   13.12106 &  0.5560  &  21.70 &  $-$18.68  &  1.27 &  0.132 & 9.0  & 1.47 & 0.60/4.4  \\
   13.12111 &  0.0890  &  22.60 &  $-$14.40  &  2.60 &  0.260 & 37.7 & 1.10 & 4.4/9.8  \\
   13.12540 &  0.4520  &  22.60 &  $-$17.61  &  1.81 &  9999 & 0   & 0.75 & 0.50/3.4  \\
   13.12545 &  0.8300  &  20.30 &  $-$21.26  &  1.69 &  0.396 & 24.0 & 0.01 & 0.60/5.0  \\
   13.12549 &  0.4930  &  21.10 &  $-$18.88  &  1.25 &  0.052 & 8.7  & 0.79 & 2.80/19.8  \\
\end{tabular}
\end{center}
\end{table}

\begin{figure}
\leavevmode
\psfig{figure=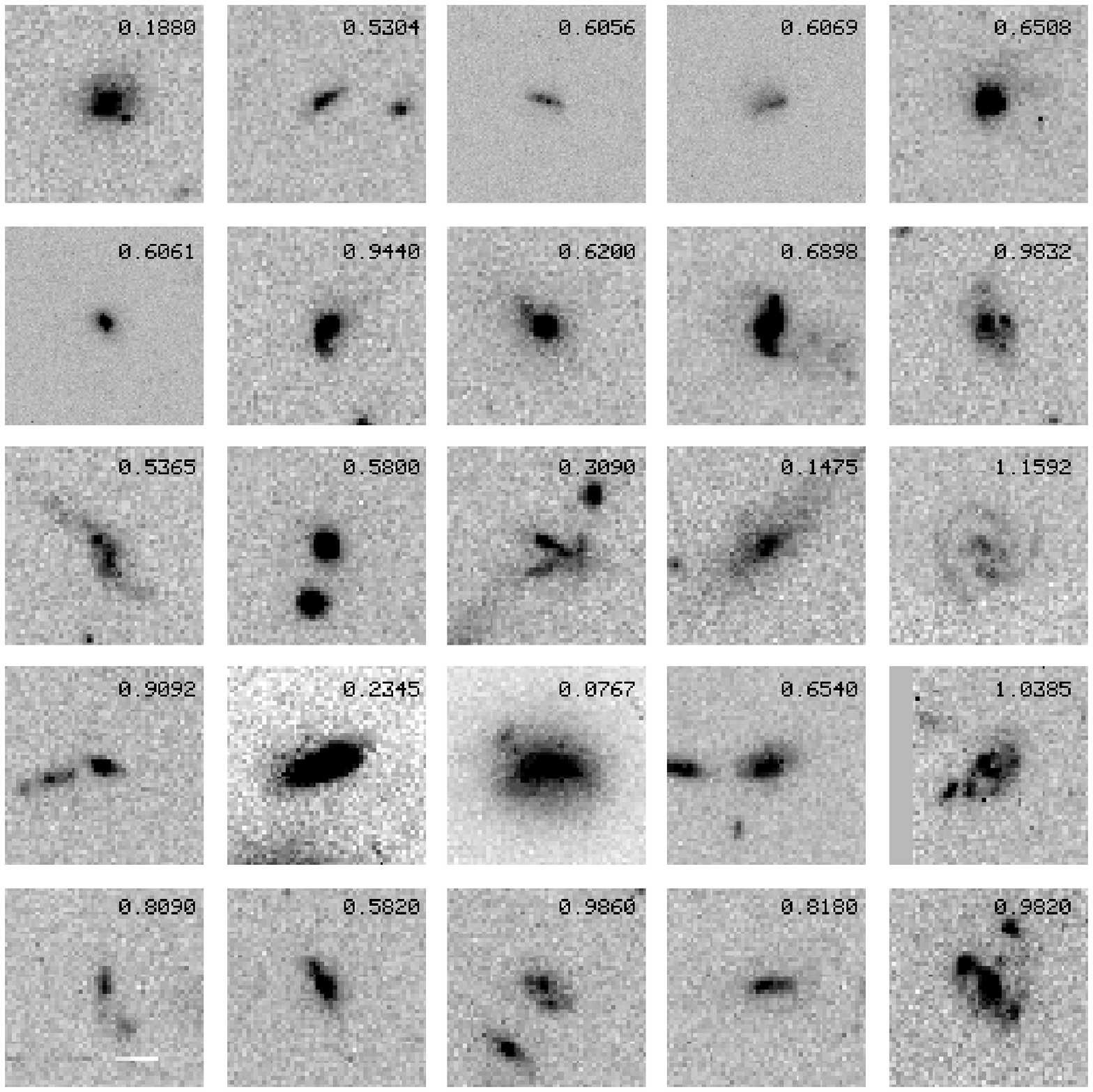,height=18cm,width=18cm}
\caption{ Galaxies classified as pairs. Each individual image is
$5\times5$ arcsec$^2$, the measured redshift is indicated on the upper
right corner of  each image.} \label{fig:pm11}
\end{figure}
\begin{figure}
\leavevmode
\psfig{figure=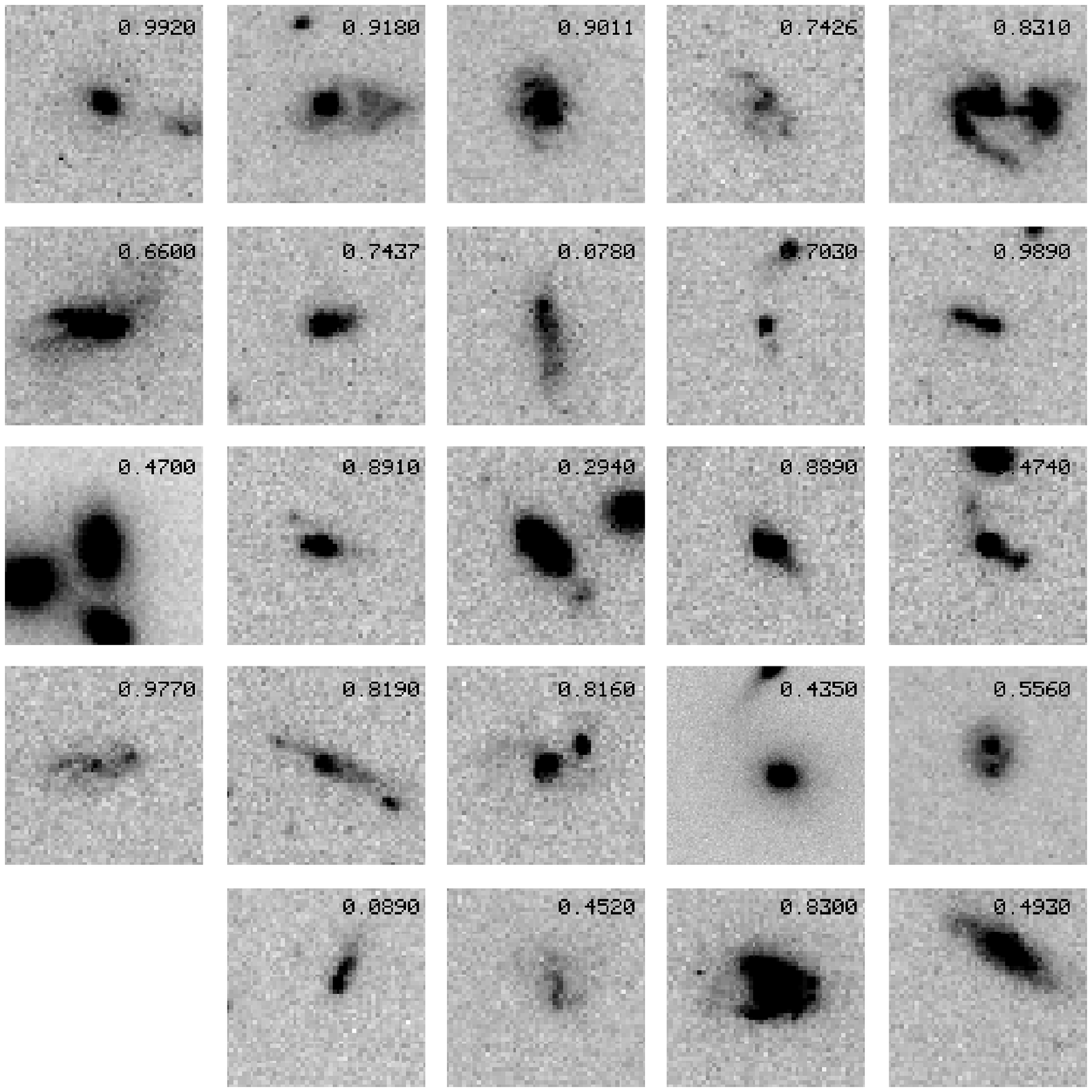,height=18cm,width=18cm}
\contcaption{ Galaxies classified as pairs. Each individual
image is $5\times5$ arcsec$^2$, the measured redshift is indicated on
the upper right corner of each image.}
\end{figure}

\begin{table} 
\begin{center}
\caption{ Pair Fraction. $N_{gal}$, $N_{maj}$ and $N_{proj}$ are the
total number of galaxies in each sample bin, the number of galaxies
classified as major mergers, and the expected background and
foreground galaxy contamination. The pair fraction is
$N_{maj}-N_{proj} / N_{gal}$; multiplied by $0.5(1+z)$ provides the
expected physical pair fraction.}
\begin{tabular}{|l|c|c|c|c|c|c|c|}
\hline
  Sample      &  z        & $N_{gal}$ & $N_{maj}$   & $N_{proj}$ &  Pair & Pair Fraction \\ 
              &           &           &             &           &   Fraction & Corr. $0.5(1+z)$ \\ \hline
All CFRS+LDSS & 0--0.2    &   40      &     6       &    22     &   0        & 0 \\
All CFRS+LDSS & 0.2--0.5  &   98      &    11       &    19     &   0        & 0 \\
All CFRS+LDSS & 0.5--0.75 &   89      &    21       &   12.2    & 9.9        & 8.1 \\
All CFRS+LDSS & 0.75--1.3 &   62      &    21       &   8.4     & 20.3       & 19.4 \\
CFRS+LDSS $M_{B_{AB}}\leq -20.5$ & 0--0.2    &    8      &     0       &   7.1     &    0   &  0 \\
CFRS+LDSS $M_{B_{AB}}\leq -20.5$ & 0.2--0.5  &   33      &     2       &   4.3     &   0    &  0 \\
CFRS+LDSS $M_{B_{AB}}\leq -20.5$ & 0.5--0.75 &   57      &    12       &   6.6     &  9.5   &  7.8 \\
CFRS+LDSS $M_{B_{AB}}\leq -20.5$ & 0.75--1.3 &   59      &    21       &   8.0     &  22.0  &  21.0 \\
\end{tabular}
\end{center}
\end{table}

\begin{table}
\begin{center}
\caption{ Rest-frame [\ion{O}{2}] equivalent width for major mergers.}
\begin{tabular}{|l|c|c|c|c|c|c|c|}
\hline
  z &  Up-coming  & $N_{gal}$ & On-going        & $N_{gal}$ &  Other &  $N_{gal}$  \\
    &  mergers   &            & mergers   &           & galaxies  &\\ \hline
$<0.5$  & 8.3    & 11 &  2.7  & 5 &  12.6  & 123 \\
0.5--0.75 & 15.0  & 7 & 37.7  & 15 & 18.9  & 67\\
$>0.75$  & 24.8 & 5 & 46.4 & 9 & 19.7 & 48 \\
\end{tabular}
\end{center}
\end{table}

\twocolumn

Our results are summarized in Table~4, which shows that the
[\ion{O}{2}] equivalent width of up-coming mergers is not
significantly different from the equivalent width of non-merging
galaxies, while the equivalent width of on-going mergers at $z\geq0.5$
is $>38$\AA, 2 times higher than for non-merger galaxies
($\sim19$\AA). On-going mergers may thus have a star formation rate
increased by a factor 2, with a corresponding luminosity increase, as
a result of the merging process. It is noteworthy that a significant
fraction of the star formation may be obscured by dust, as dust is
often seen in merging systems, and therefore, that the star formation
rate increase measured from [\ion{O}{2}] should be taken as a lower limit
(Hammer et al. 1998).

\section{Bias in ground based selected samples from merger induced
luminosity enhancement}\label{s3}

In this section we examine the effect of having selected our
galaxy sample from a redshift survey based on ground-based
imaging. Some of the galaxies in the redshift survey are now
identified from HST imaging as pairs of nearby galaxies unresolved
in ground-based data. Seeing effects clearly lead ground-based
surveys to include some galaxies in magnitude limited samples only
because the added luminosity from a close companion is sufficient
to produce a total blended luminosity above a limiting magnitude
threshold. From the analysis of the magnitude difference with the
closest companion as described in Section~\ref{le}, we find that in the
galaxies of the CFRS sample imaged with HST, 11 (of 232) galaxies
have been included in the ground-based redshift survey because of
this effect. This is most significant for $z>0.7$, as 6 (of 73)
galaxies above this redshift are in the complete HST-CFRS sample
because of this effect, which enters at the $8\pm3$\% level. This
bias also demonstrates the difficulties inherent in comparing deep
galaxy counts obtained from ground-based and space-based
observations.

Taking this effect into account would lower the physical pair
fraction, corrected for background/foreground contamination, from
9.9\% to 8.1\% at $z=0.63$ and from 20.3\% to 14.5\% at $z=0.91$. This
would lower the parameter $m$ describing the physical pair rate
change with redshift to $m=2.7 \pm 0.6$.

\section{Discussion and Conclusion}

The visual classification of mergers, and the pair count analysis
performed using manual classifications and using the Lee
classifier, paint a consistent picture of the effects of mergers
on high redshift galaxies. Our analysis shows that major mergers
of galaxies ($\delta m \leq 1.5$) are playing an increasingly
important role at higher redshifts, both in terms of number evolution
and in inducing a significant luminosity enhancement. From the
visual classification of mergers, we find that the merger fraction
evolves as $(1+z)^{m}$, with $m=3.4\pm0.6$, in excellent agreement
with the merger fraction derived from the pair counts, for which
$m=3.2\pm0.6$.

As shown explicitly in Section~\ref{s3}, ground-based
magnitude-limited imaging samples can be biased by the effects of
seeing, as close pairs of galaxies are counted singly. Around
$\sim8$\% of high redshift $z > 0.7$ galaxies that are formally
below the selection limit would be included in a ground based
$I_{AB} \leq 22.5$ mag sample due to this effect. If our magnitude
limited redshift survey had been defined from HST imaging data
rather than from imaging data limited by ground based seeing, the
pair fraction at $z \sim 0.9$ would be lowered from $\sim20\%$ to
$14.5\%$. After correction for this bias, we conclude that the
merger fraction evolution exponent can best be parameterized as
$m=2.7\pm0.6$.

Transforming the observed merger fraction into a merger rate
requires knowledge of the average timescale over which a merger is
completed (i.e. over which morphological traces of the merger are
gone at the 1~kpc resolution at which we are observing the present
sample). Although there are considerable uncertainties in
estimating this timescale, an upper limit of 400~Myr to 1~Gyr
seems reasonable from both data and simulations (e.g. Patton et
al. 1997; Mihos \& Hernquist 1994b). Our results indicate that
between redshift 0 and 0.9, the merger probability is $\sim
0.38\times \frac{2 Gyr \times q_0^{0.5}}{\tau}$, where $\tau$ is
the merger timescale, and $\sim 0.2\times \frac{2 Gyr \times
q_0^{0.5}}{\tau}$ between $z=0.75$ and $z=1.2$. Using a (conservative)
range of merger timescales from 400~Myr to 1~Gyr, these relations
suggest a galaxy will undergo on average 0.8 to 1.8 merger events
from $z=1$ to $z=0$, with 0.5 to 1.2 merger events occurring in a
2~Gyr time span at $z\sim0.9$.

Semi-analytical models of galaxy formation in hierarchical
clustering theories make strong predictions regarding the
high-redshift merger rate. The model of Baugh et al. (1996)
predicts that more than 50\% of the elliptical galaxies, and 15\%
of the spiral galaxies suffered a major merger event in the
redshift interval $0.0 \leq z \leq 0.5$, while by $z \sim 1$,
these numbers increase to more than 90\% and 55\% for ellipticals
and spirals respectively, indicating the lower and upper limits
for the whole population of galaxies (Baugh et al. 1996). These
values appear broadly consistent with the numbers derived above
from our data.

As the galaxies merge, two galaxies will be replaced by one which
will be on average 0.5 magnitudes brighter from the simple
co-addition of the initial luminosities, and an additional
temporary star formation rate increase of a factor of $\sim$2,
triggered by the merger event. With the increase in the number of
mergers at high redshifts, merger events will contribute
significantly to the evolution of the luminosity function by
making L$^{*}$ brighter, and the number density smaller, with changing
redshift. The analysis we have conducted has mainly been targeted
at the most obvious merger events. We note that the number
evolution and luminosity evolution that we have derived probably
minimizes the full effect of mergers as the image resolution and
depth of our HST data do not allow us to take into account mergers
of smaller galaxies with bright galaxies in our sample (i.e.
systems with $\delta m > 1.5$ mag).

Our results  demonstrate the importance of mergers in the
evolution of the luminosity function and of the luminosity density
of the universe out to $z \sim 1 $. The derivation of the merger
fraction and of the merger rate at still higher redshifts is of
great interest in order to determine the contribution of mergers
to the putative peak or flattening in the UV luminosity density of the universe
inferred to lie at a redshift between 1 and 2 (Madau et al. 1996).

\bsp 
\label{lastpage}
\end{document}